\documentclass[letter,twocolumn]{jpsj2}
\usepackage{bm}
\usepackage{graphicx}
\usepackage{amsmath,amsfonts,amsthm,amssymb}

\def\eqlt{\mathrel{\mathpalette\@vereq<}}  
\def\eqgt{\mathrel{\mathpalette\@vereq>}}  
\def\@vereq#1#2{\lower2.5pt\vbox{\baselineskip0pt \lineskip-.5pt
  \ialign{$\m@th#1\hfil##\hfil$\crcr#2\crcr{=}\crcr}}}

\def\runauthor{Aimi and Imada}

\title{%
Does Simple Two-Dimensional Hubbard Model Account for High-$T_c$ Superconductivity in Copper Oxides?
}

\author{%
Takeshi Aimi and Masatoshi Imada$^1$
}

\inst{%
Department of Physics, University of Tokyo, 7-3-1 Hongo, Bunkyo-ku, Tokyo 113-0033\\
$^1$Department of Applied Physics, University of Tokyo, 7-3-1 Hongo, Bunkyo-ku, Tokyo 113-8656
}

\abst{
We reexamine whether the essence of high-$T_c$ superconductivity is contained in doped Hubbard models on the square lattice by using recently developed pre-projected Gaussian-basis Monte Carlo method. The superconducting correlations of the $d_{x^2-y^2}$ wave symmetry in the ground state at distance $r$ decays essentially as $r^{-3}$.  The upper bound of the correlation at long distances estimated by this unbiased method is $10^{-3}$, indicating that recent extensions of dynamical mean-field theories and variational methods yielded at least an order of magnitude overestimates of it. The correlations are too weak for the realistic account of the cuprate high-$T_c$ superconductivity.
}
\kword{%
Monte Carlo method, doped Hubbard model, $d_{x^2-y^2}$-wave superconductivity, cuprate superconductor, high-temperature superconductivity
}
\begin{document}

\maketitle

Since the proposal that the simple Hubbard model on a square lattice 
captures essential physics of the high-temperature
copper-oxide superconductors~\cite{Anderson}, the model has been extensively studied
and now become one of the most intensively studied issues in condensed matter physics. 
However, the ground state of this model, particularly whether the high-$T_c$
superconductivity 
is accounted for, has long been highly controvertial.

The Hubbard Hamiltonian contains the on-site Coulomb interaction $U$ and the hopping integral $t$ as
\begin{eqnarray}
\hat{H}=-t\sum_{\langle i,j\rangle,\sigma}^{N_s}(\hat{c}_{i\sigma}^{\dag}\hat{c}_{j\sigma}+h.c.)+U\sum_{i}^{N_s}\hat{c}_{i\uparrow}^{\dag}\hat{c}_{i\uparrow}\hat{c}_{i\downarrow}^{\dag}\hat{c}_{i\downarrow}, 
\label{eq:hamiltonian} 
\end{eqnarray}
where $\hat{c}_{i\sigma}^{\dag}(\hat{c}_{i\sigma})$ is
the creation (annihilation) operator of an electron with spin $\sigma$ on the $i$-th site for $N_{s}$-site lattice and $\langle i,j\rangle$ represents nearest neighbor pairs. 
 
In the weak to intermediate coupling region, 
the auxiliary-field quantum Monte Carlo (AFQMC) method gave negative results for the 
superconductivity around up to $U/t=4$ by the analysis of the 
superconducting correlation functions~\cite{Imada,Furukawa}. 
Zhang {\it et al.} calculated pairing correlations of 
the $d_{x^2-y^2}$-wave channels  
by combining the  
constrained-path approximation (CPMC) and concluded the absence of the 
superconductivity~\cite{Zhang}. 
Due to the negative sign problem known as a major obstacle of the Fermion simulation methods, 
the AFQMC method does not offer converged results for larger $U/t$.

On the other hand, with the dynamical cluster approximation (DCA), Maier {\it et al.} 
claimed, from the cluster sizes up to 26 sites, 
that the $d_{x^2-y^2}$-wave pair-field susceptibility indicates 
the transition to the $d_{x^2-y^2}$-wave superconductor 
at the critical temperature $T_{c}\approx 0.023t$ at 10\% doping, even at $U/t=4$~\cite{Maier}. 
Cellular dynamical mean-field theory (CDMFT), cluster perturbation theory and variational cluster theory (VCT) gave similar results~\cite{Capone, Tremblay, Potthoff}.
Accuracy of the mean-field treatment contained in these studies remains to be examined by larger clusters. 
Numerical results based on variational Monte Carlo (VMC) methods suggest the firm $d_{x^2-y^2}$-wave superconductivity for $U/t\gtrsim 6$
upon doping~\cite{Yokoyama3}.
However, the VMC method could overestimate the stability of superconductivity by relatively worse estimate of energy for correlated metals. 
From these controversies, one finds that
the stability of superconductivity 
in Hubbard models still remains a fundamental open issue. 
In the circumstance of the ground state highly competing
with others, unbiased accurate approaches are needed to answer whether the realistic amplitude of the copper-oxide superconductivity is accounted for. 

In this letter, we reanalyze the pairing correlations of the $d_{x^2-y^2}$ 
symmetry on doped Hubbard models by using recently developed
pre-projected Gaussian-basis Monte Carlo (PR-GBMC) method~\cite{Aimi}. 
Thanks to the absence of the negaitive sign problem, 
this method allows us to go beyond the tractable range of 
the conventional methods including doped and larger $U$ region without any approximation. 
We reveal that the pairing correlations do not show 
distinct enhancement up to $U/t=7$. 
An accurate estimate of the upper bound of 
the superconducting correlation ($\sim 10^{-3}$) is far below the recent results by DCA, CDMFT, VMC and VCT, indicating the necessity of 
the higher accuracy than these studies in the literature for assessing the possible superconductivity in the Hubbard models. 


Our PR-GBMC method is described in detail and extensively benchmarked elsewhere~\cite{Aimi}. 
Here we only explain the basic procedure of PR-GBMC. 
To calculate the ground state of the Hamiltonian (\ref{eq:hamiltonian}), 
we solve the Liouville equation of the density-matrix operator $\hat{\rho}(\tau)$ 
(here, $\tau$ denotes the imaginary time): 
\begin{align}
\frac{\partial\hat{\rho}(\tau)}{\partial\tau}=-\frac{1}{2}\left[\hat{H},\hat{\rho}(\tau)\right]_{+} \label{eq:Liouville}
\end{align}
using an expansion of $\hat{\rho}(\tau)$ 
by a Gaussian-basis $\hat{\Lambda}(\Omega,\bm{n})$: 
\begin{align}
\hat{\rho}(\tau)&=\int d\Omega d\bm{n}P(\Omega,\bm{n};\tau)\Lambda(\Omega,\bm{n}), \label{eq:expansion} \\
\hat{\Lambda}(\Omega,\bm{n})&=\Omega\det(\bm{I}-\bm{n}):e^{-\hat{\bm{c}}^{\dag}[2\bm{I}+(\bm{n}^{T}-\bm{I})^{-1}]\hat{\bm{c}}}: , 
\end{align}
where $P(\Omega,\bm{n};\tau)$ is expansion coefficient, 
$\bm{I}$ is a $2N_{s}\times 2N_{s}$ 
unit matrix and the vector operator $\hat{\bm{c}}^{\dag}$ is defined by 
\begin{align}
\hat{\bm{c}}^{\dag}=(\hat{c}_{1\uparrow}^{\dag},\hat{c}_{2\uparrow}^{\dag},\cdots,\hat{c}_{N_{s}\uparrow}^{\dag},\hat{c}_{1\downarrow}^{\dag},\cdots,\hat{c}_{N_{s}\downarrow}^{\dag}) .
\end{align}
The colon bracket $:\quad:$ is the normal ordering operator. 
One of the parameters of the Gaussian-basis 
$\Omega$ works as the weight of the importance sampling of the 
Monte Carlo procedure 
and the other parameter 
$\bm{n}$ is a $2N_{s}\times 2N_{s}$ matrix characterized 
by 
$n_{(i\sigma),(j\sigma)}={\rm Tr}\left[\hat{c}_{i\sigma}^{\dag}\hat{c}_{j\sigma}\hat{\Lambda}\right]/{\rm Tr}\left[\hat{\Lambda}\right]
={\rm Tr}\left[\hat{c}_{i\sigma}^{\dag}\hat{c}_{j\sigma}\hat{\Lambda}\right]/\Omega$. 
By using the expansion (\ref{eq:expansion}), 
the Liouville equation (\ref{eq:Liouville}) is 
mapped into a Fokker-Planck equation for $P$.
It should be noted that $P$ can be taken positive definite~\cite{CorneyUnpub,Aimi}.
In the actual calculation, 
we solve the corresponding Langevin equations 
with respect to $\Omega$ and $\bm{n}$~\cite{Corney1}. 
In our PR-GBMC method~\cite{Aimi}, by using quantum-number projectors~\cite{Assaad}
$\hat{P}=\int d\bm{x}g(\bm{x})e^{i\hat{\bm{c}}^{\dag}\bm{h}(\bm{x})\hat{\bm{c}}}$, 
the weight $\Omega$ is transformed into a quantum-number-projected weight $\tilde{\Omega}$. 
Here, $\bm{h}(\bm{x})$ is a $2N_{s}\times 2N_{s}$ Hermitian matrix and $g(\bm{x})$ 
is an integration weight with $\bm{x}$ being a parameter of the quantum-number projection 
such as phase, Euler angles of the spin-space, {\it etc}., . 
The importance sampling is performed by 
the quantum-number-projected weight $\tilde{\Omega}$ defined by 
\begin{equation}
\tilde{\Omega}=\Omega\int d\bm{x}g(\bm{x})\det\left[(e^{i\bm{h}(\bm{x})}-\bm{I})\bm{n}^{T}+\bm{I}\right]. 
\end{equation}
We have taken large enough $\tau$ to assure the convergence to the ground state and confirmed the absence of the boundary term in $\hat{P}$ following the procedure in Ref.~\cite{Aimi}. 

To reduce finite-size effects and estimate it in a systematic 
fashion, we mainly employ fillings at which the corresponding noninteracting systems show closed-shell structure~\cite{Furukawa}. 
For comparison, open-shell conditions are also studied in some cases. 

The accuracy of the present method is seen typically in the estimate of the energy per site for the Hubbard model at half filling~\cite{Aimi}: $-0.8575\pm 0.0005$ for $6\times 6$ lattice in comparison with $-0.8575\pm 0.0008$ by the accurate AFQMC result and  $-0.8595\pm 0.0005$ for $8\times 8$ lattice in comparison with $-0.8607\pm 0.0009$ by the AFQMC result. This accuracy is more than one order of magnitude better than typical VMC accuracy.

We now present results for the equal-time pairing correlation defined as 
\begin{eqnarray}
P_{\alpha}(r)=\frac{1}{2N}\sum_{i=1}^{N}\langle\Delta_{\alpha}^{\dag}(i)\Delta_{\alpha}(i+r)+\Delta_{\alpha}(i)\Delta_{\alpha}^{\dag}(i+r)\rangle , 
\end{eqnarray}
where $\Delta_{\alpha}$ is the 
 order parameter defined as 
\begin{equation}
\Delta_{\alpha}(i)=\frac{1}{\sqrt{2}}\sum_{r}f_{\alpha}(r)(\hat{c}_{i\uparrow}\hat{c}_{i+r\downarrow}-\hat{c}_{i\downarrow}\hat{c}_{i+r\uparrow}). 
\label{Delta.eq}
\end{equation}
Here, $f_{\alpha}(r)$ is the form factor defined as
\begin{eqnarray}
f_{2d}(r)=\delta_{r_{y},0}(\delta_{r_{x},1}+\delta_{r_{x},-1}) 
-\delta_{r_{x},0}(\delta_{r_{y},1}+\delta_{r_{y},-1})
\label{formfactor.eq}
\end{eqnarray}
with $\delta_{ij}$ being Cronecker's delta. 
The suffix $\alpha=2d$ represents 
the $d_{x^{2}-y^{2}}$-wave. 
In addition to $P_{\alpha}(r)$, we also calculate the superconducting correlation 
defined by the summation of 
$P_{\alpha}(r)$ over $r$: 
\begin{align}
S_{\alpha}=\frac{1}{4}\sum_{r}P_{\alpha}(r) , 
\end{align}
where we multiply the factor $1/4$ 
to allow direct comparison with the data 
in Refs.~\citen{Furukawa} and~\citen{Furukawa2}. 

\begin{figure}[h]
\begin{center}
\includegraphics[width=5.4cm]{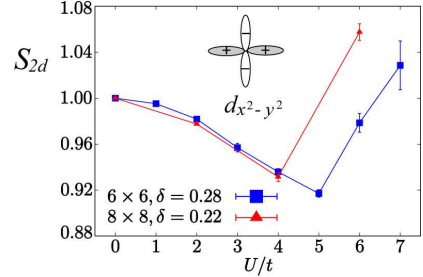} \\
\caption{(Color online) Pairing correlation for $d_{x^2-y^2}$-wave as function of on-site interaction $U/t$. 
Squares and triangles represent 
$6\times 6$ lattice at the doping concentration $\delta=0.28$ and 
$8\times 8$ lattice at $\delta=0.22$, respectively. 
} \label{fig:super_sum2}
\end{center}
\end{figure}
\begin{figure}[h]
\begin{center}
\includegraphics[width=6cm]{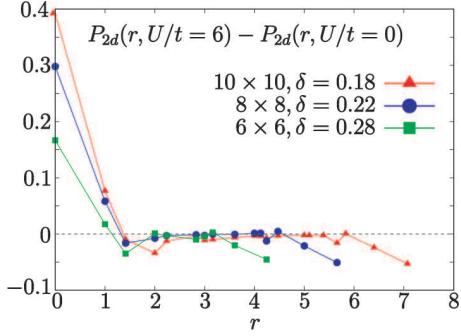}
\caption{(Color online) Spatial dependence of pairing correlations of the
$d_{x^2-y^2}$-wave at $U/t=6$ subtracted by 
those at $U/t=0$. 
The triangles, circles and the squares represent 
sizes
$6\times 6$ lattice at the doping concentration $\delta=0.28$, 
$8\times 8$ lattice at $\delta=0.22$ and 
$10\times 10$ lattice at $\delta=0.18$, respectively. 
} \label{fig:sabun}
\end{center}
\end{figure}

Figure \ref{fig:super_sum2} shows that
$S_{2d}$ decreases first, then increases above $U/t\sim 5$
with the increase of $U$.
The sharp enhancement of $S_{2d}$ seems to be consistent with 
the enhancement of the condensation energy above $U/t\sim6$ seen in the VMC results 
(see Fig.3 in Ref.~\citen{Yokoyama3}). 
The most of the enhancement observed in $S_{2d}$, however,
comes from the short-range correlations. 
Figure \ref{fig:sabun} shows the pairing correlation 
of the $d_{x^2-y^2}$-wave at $U/t=6$ 
subtracted by that at $U/t=0$. 
It shows that the on-site and nearest-neighbor 
pair correlations exhaust most of the enhancement in $S_{2d}$. 
The enhancement in the short-range part was also pointed out by the CPMC results
\cite{Zhang}. 

The enhancement of the on-site correlation $P_{2d}(r=0)$ comes from 
the nearest-neighbor spin and charge correlations. 
In fact, $P_{2d}(r=0)$ contains the sum over the nearest-neighbor spin and charge correlations, $C=\langle -4\bm{S}_{0}\!\cdot\!\bm{S}_{\hat{x}}+N_{0}N_{\hat{x}}\rangle$,
where $\hat{x}$ denote the unit vectors in the $x$ directions. 
$\bm{S}$ and $N$ are spin and charge operators, respectively. 
The enhancement is well accounted for by the enhancement of $C$, 
which is not related to the superconductivity but to enhanced antiferromagnetic correlations. 

\begin{figure}[h]
\begin{center}
\includegraphics[width=6.5cm]{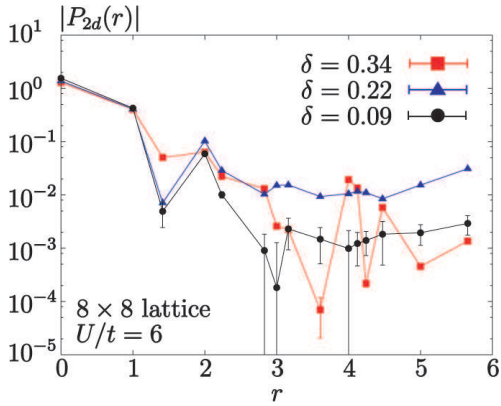}
\caption{(Color online) Filling dependence of pairing correlation $P_{2d}(r)$ as function of distance $r$ on the
$8\times 8$ lattice at $U/t=6$. 
Squares, triangles and circles represent $P_{2d}(r)$ 
at $\delta=0.34$, $0.22$ and $0.09$, respectively. 
} \label{fig:8x8n}
\end{center}
\end{figure}

Next, we analyze the filling dependence of the $d_{x^2-y^2}$-wave 
pairing correlation $P_{2d}(r)$. 
As we see in Fig.\ref{fig:8x8n}, 
the pairing correlation $P_{2d}(r)$ has the largest values at the doping $\delta=0.22$. 
Although $\delta=0.22$ satisfies the closed shell condition, 
the other two fillings are under the open shell condition. 
We do not find an appreciable difference between open and closed shell conditions. 

To gain insight into the main issue whether the non-zero offset 
exists in the long-range part of the pairing correlation, 
we analyze system size dependence of $P_{2d}(r)$ 
at the filling around $\delta\sim 0.2$, where the long-range part of 
$P_{2d}(r)$ may have the largest values. 
\begin{figure}[h]
\begin{center}
\includegraphics[width=7.5cm]{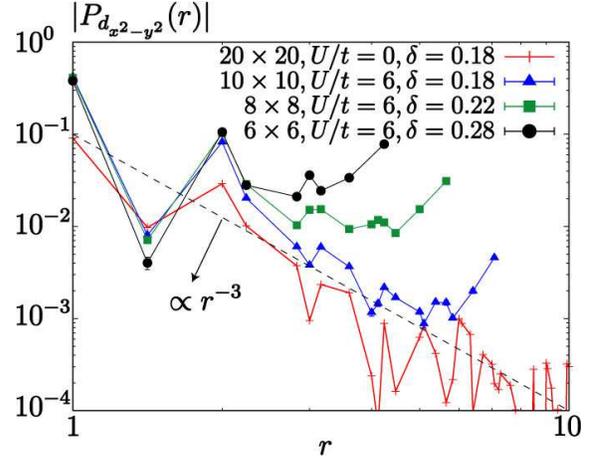}
\caption{(Color online) System size dependence of pairing correlations $P_{2d}(r)$ 
at $U/t=6$ and $\delta\sim 0.2$. 
Triangles, squares and circles represent the
$10\times 10$, $8\times 8$ and $6\times 6$ lattices. 
The pairing correlations at $U=0$ are illustrated by 
the crosses (for $20\times 20$ lattice). The dashed line represents the 
asymptotic $r^{-3}$ scaling for the noninteracting system. 
} \label{fig:size_2d}
\end{center}
\end{figure}
Figure \ref{fig:size_2d} shows $P_{2d}(r)$ 
for several system sizes, where $P_{2d}(r)$ 
decreases with the increasing system size as well as with the increasing distance $r$ at large $r$. 
It is known that in the non-interacting case, the pairing correlation decays asymptotically as 
$P_{2d}(r)\propto r^{-3}$ as is inferred from 20$\times$20 lattice result in Fig.\ref{fig:size_2d}~\cite{Furukawa}. 
This behavior is 
gradually reproduced with the increase of the system size even at $U/t=6$. 
It should be noted that in addition to the power-law decays, the correlations show increase again at the farthest distances in finite size systems
({\it e.g.}, $r\ge 3$ for $6\times 6$, $r\ge 4$ for $8\times 8$ and $r\ge 5$ for $10\times 10$ lattices in Fig.\ref{fig:size_2d}). 
These increases are well 
accounted for by the superimpose of the tail from the 
other directions due to the periodic boundary condition. 
After correcting this artifact, the pairing correlation 
shows no signal of level off to a nonzero correlation at long distances. 
From converged values of correlations up to $r\sim 5$, we may safely conclude that 
$|P_{2d}(r\rightarrow\infty)|\leq 10^{-3}$ 
for the Hubbard model at $U/t=6$. This main result of this paper poses a severe constraint on the possible superconductivity.

When we could assume the existence of the $d_{x^2-y^2}$-wave 
superconducting long-range order,
the order parameter average $\langle \Delta_{2d} \rangle$ should be related to $P_{2d}$ as 
$|P_{2d}(r\rightarrow \infty)|\simeq \langle \Delta_{2d} \rangle^2$.
Although recent DCA\cite{Maier}, CDMFT\cite{Capone}, VCT\cite{Potthoff} and VMC\cite{Yokoyama3} studies are not necessarily consistent each other, a common typical result in these studies is that $\langle \Delta_{2d} \rangle$ in the corresponding definition by Eq.(\ref{Delta.eq}) has the amplitude around $0.1$\cite{Capone} or even larger\cite{Potthoff}, which leads to $|P_{2d}(r\rightarrow\infty)|\gtrsim 0.01$. In fact, the VMC study\cite{Yokoyama3} indicates $|P_{2d}(r\rightarrow\infty)|\simeq 0.01-0.02$ consistently. However, these are one order of magnitude larger than our upper bound.  

The overestimate of the order parameter by the DCA, CDMFT, VCT and VMC may be ascribed to the enhancement of superconducting correlations found only in the short-ranged part in the present study. In fact, it is natural to expect that the enhancement in the short-ranged part generates overestimated pairing in the mean-field type treatment or in the small cluster studies in which long-ranged part of the fluctuations is ignored.  It would be desired to evaluate the long-range part of correlations in extensions of the dynamical mean field theory to make the size scaling to the thermodynamic limit possible and take into account long-range fluctuation effects in the VMC studies.

The superconducting order parameter $\Delta_{2d}$ has close connection to the estimate of other physical quantities such as the condensation energy $Q$, the single-particle gap $\tilde{\Delta}_{2d}$ and the superconducting transition temperature $T_c$.  If $|P_{2d}|$ would be one order of magnitude reduced, since $Q$ may be scaled by $|P_{2d}(r\rightarrow\infty)|=\Delta_{2d}^2$, the condensation energy obtained as $Q\simeq 10^{-3}t$\cite{Yokoyama3,Yamaji} should be reduced to $\simeq 10^{-4}t$.  This upperbound of $Q$ with $t\sim 0.4$ eV, which is a realistic value of the cuprate superconductors, results in $Q < 0.04$ meV. This is at least one order of magnitude smaller than the experimental results ($\sim 0.2-0.4$meV)~\cite{Hao,Triscone,Loram}.

The single-particle gap amplitude $\tilde{\Delta}_{2d}$ in the antinodal direction 
has been estimated by CDMFT to be around 0.1$t$\cite{Kancharla}. However, this gap is scaled linearly by the superconducting order parameter $\Delta_{2d}$. When the overestimate by the factor 3 for $\Delta_{2d}$ is corrected, we find $\tilde{\Delta}_{2d}\simeq 0.03t$ as an upperbound. 
If we again take $t\sim 0.4$ eV and employ the experimentally observed gap amplitude 
$\tilde{\Delta}_{2d}\sim 40$-$50$ meV~\cite{Tanaka}, the experimental value of $\tilde{\Delta}_{2d}\simeq 0.1 t$ is at least factor 3 larger.

Maier {\it et al.} claimed that from the DCA results, 
the transition to the $d_{x^2-y^2}$-wave superconductor 
occurs at $T_{c}\approx 0.023t$ at $\delta=0.1$, even at $U/t=4$~\cite{Maier}. 
(Note that ``$T_c$" should be regarded as the critical temperature of 
the Beresinski-Kosterlitz-Thouless transition if a strictly two-dimensional 
system is considered, whereas it is known that the additional three dimensionality would not change $T_c$ much.) Since $T_c$ is scaled by the gap amplitude, it may again be scaled by $\sqrt{|P_{2d}(r\rightarrow\infty)|}$ and $T_c$ may be overestimated by DCA at least by a factor 3.  Again by taking $t\simeq 0.4$eV, the upperbound of $T_c$ may be 30 K, which does not offer a realistic model for the cuprate high-$T_c$ superconductors with $T_c>100$K.

For the moment, it is not known whether this restriction could be 
relaxed for larger $U$ beyond $7t$.
At least in the intermediate coupling region with $U$ 
comparable to the bare bandwidth, 
which is realistic for the cuprate superconductors, 
the Hubbard model does not seem to offer a realistic 
account of the superconductivity in the right order of amplitude.
The form factor $f$ spatially more extended than Eq.(\ref{formfactor.eq}) does not change these difficulties.

Effects of reduced quasiparticle weight on the pairing~\cite{Dagotto} is an unexplored and important issue to be clarified for theoretical prediction on the relation between $P_{2d}$ and the single-particle gap as well as $T_c$ and $Q$. However, it does not alter our conclusion that the available results on the Hubbard model mentioned above~\cite{Maier,Capone,Tremblay,Potthoff,Yokoyama3} and $t$-$J$ models~\cite{Randeria,Sorella} obtained by calculating the same order parameters from variational or mean-field approximations do not provide a clue for understanding the high $T_c$ superconductivity because of the substantial overestimates of the pairing correlations in their approaches. 
Accuracy of numerical approaches higher than that obtained by these methods is required.

 
We have calculated the superconducting correlations of 
the doped Hubbard model by using the PR-GBMC method. 
The present result poses constraints on the occurence of the superconductivity in the 
Hubbard model on the square lattice.  Up to $U/t=7$, the $d$-wave correlation is smaller than $10^{-3}$ at long distances, which is much smaller than the recent approximate estimates by the variational Monte Carlo methods and extensions of the dynamical mean field theory. Comparing with available theoretical works we estimated the upper bounds for the amplitudes of the superconducting gap, condensation energy and $T_c$.  The present estimates of the upper bounds are far below those of the copper oxide superconductors.  Thus our conclusion is that the simple Hubbard model does not offer a model for the superconductivity in the right order of amplitude.  For further progress, we need an accurate estimate of the correlations at long distances by fully taking account of fluctuations.
Our results show that the short-ranged Coulomb repulsion by itself does not automatically
guarantee the emergence of the superconductivity at high critical temperatures
even when it is close to the antiferromagnetic order in two dimensional systems
or in the close proximity to the Mott insulator.  This is consistent with the fact that many doped Mott insulators and metals near the antiferromagnetic quantum critical point do not automatically show the superconductivity even when the residual resistivity is very low~\cite{RMP}. We are urged to examine more detailed conditions beyond the simple framework of the Hubbard model, such as the internal structure of the Mott criticality~\cite{imada2006} to understand the cuprate superconductors. 

A part of our computation has been done at 
the supercomputer center at ISSP, Univ. of Tokyo. 
This work is partially supported by a Grant-in-Aid 
under the grant numbers 
17071003 and 16076212 from MEXT, Japan.


\begin{thebibliography}{9}

\bibitem{Anderson} 
P. W. Anderson: Science {\bf 235} (1987) 1196. 

\bibitem{Imada}
M. Imada: J. Phys. Soc. Jpn. {\bf 60} (1991) 2740.

\bibitem{Furukawa}
N. Furukawa, and M. Imada: J. Phys. Soc. Jpn. {\bf 61} (1992) 3331. 

\bibitem{Zhang}
S. Zhang, J. Carlson, and J. E. Gubernatis: Phys. Rev. Lett. {\bf 78} (1997) 4486. 

\bibitem{Maier} 
T. A. Maier, M. Jarrell, T. C. Schulthess P. R. C. Kent, and J. B. White: Phys. Rev. Lett. {\bf 95} (2005)  
237001. 

\bibitem{Tremblay}
D. S\'en\'echal, P.-L. Lavertu, M.-A. Marois and A.-M.S. Tremblay:
Phys. Rev. Lett. {\bf 94} (2005) 156404.

\bibitem{Capone} 
M. Capone and G. Kotliar: Phys. Rev. B {\bf 74} (2006) 054513.

\bibitem{Potthoff}
M. Aichhorn, E. Arrigoni, M.  Potthoff, and W. Hanke: Phys. Rev. B {\bf 74} (2006) 024508.

\bibitem{Yokoyama3} 
H. Yokoyama, M. Ogata and Y. Tanaka:
J. Phys. Soc. Jpn. {\bf 75} (2006) 114706; 
H. Yokoyama, Y. Tanaka, M. Ogata, and H. Tsuchiura: 
 ibid. {\bf 73} (2004) 1119. 

\bibitem{Aimi} 
T. Aimi, and M. Imada: J. Phys. Soc. Jpn. {\bf 76} (2007) 084709.

\bibitem{CorneyUnpub} 
J. F. Corney, unpublished.

\bibitem{Corney1} 
J. F. Corney, and P. D. Drummond: Phys. Rev. B {\bf 73} (2006) 125112; 
J. Phys. A: Math. Gen. {\bf 39} (2006) 269.

\bibitem{Assaad}
F. F. Assaad, P. Werner, P. Corboz, E. Gull, and M. Troyer:
Phys. Rev. B {\bf 72} (2005) 224518.

\bibitem{Furukawa2}
Equation (5.1) in Ref.~\cite{Furukawa} should be corrected as 
$S_{{\rm sup}}^{\alpha}=\frac{1}{4}\frac{1}{2N_{s}}\sum_{ij}\langle\Delta_{\alpha}^{\dag}(i)\Delta_{\alpha}(j)+\Delta_{\alpha}(i)\Delta_{\alpha}^{\dag}(j)\rangle$
so as to be directly compared with Fig.19 in Ref.~\cite{Furukawa}. 

\bibitem{Yamaji}
K. Yamaji, T. Yanagisawa and M. Miyazaki: Physica C {\bf 445-448} (2006) 171.

\bibitem{Hao} 
Z. Hao, J.R. Clem, M.W. McElfresh, L. Civale, A.P. Malozemoff, and F. Holtzberg:
Phys. Rev. B (1991) {\bf 43}, 2844.

\bibitem{Triscone} 
G. Triscone, B. Revaz, A.F. Khoder, J.-Y. Genoud, A. Junod, and J. Muller:
Physica C {\bf 235-240} (1994) 1557.

\bibitem{Loram} 
J.W. Loram, K.A. Mirza, J.R. Cooper, and W.Y. Liang: Phys. Rev.Lett. {\bf 71} (1993) 1740; 
J.W. Loram, J.R. Cooper and J. Tallon: Physica C {\bf 341-348} (2000) 1837.
 
\bibitem{Kancharla} 
S.S. Kancharla {\it et al.}: cond-mat/0508205.

\bibitem{Tanaka}
K. Tanaka, W.S. Lee, D.H. Lu, A. Fujimori, T. Fujii, Risdiana, I. Terasaki, D.J. Scalapino, T.P. Devereaux, Z. Hussain, and Z.-X. Shen: 
Science {\bf 314} (2006) 1910;
H. Ding, J.R. Engelbrecht, Z. Wang, J.C. Campuzano, S.-C. Wang, H.-B. Yang, R. Rogan, T. Takahashi, K. Kadowaki, and D.G. Hinks, 
Phys. Rev. Lett. {\bf 87} (2001) 227001.


\bibitem{Dagotto}
E. Dagotto and J.R. Schrieffer: Phys. Rev. B {\bf 43} (1991) 8705. 

\bibitem{Randeria}
A. Paramekanti, M. Randeria and N. Trivedi:
Phys. Rev. Lett. {\bf 87} (2001) 217002.

\bibitem{Sorella}
S. Sorella, G.B. Martins, F. Becca, C. Gazza, L. Capriotti, A. Parola and E. Dagotto:
Phys. Rev. Lett. {\bf 88} (2002) 117002.

\bibitem{imada2006}
M. Imada, Phys. Rev. B: {\bf 72} (2005) 075113; T. Misawa and M. Imada, ibid. {\bf 75} (2007) 115121.

\bibitem{RMP}
M. Imada, A. Fujimori and Y. Tokura: Rev. Mod. Phys. {\bf 70} (1998) 1039.
\end{thebibliography}
\end{document}